\documentclass[twocolumn,secnumarabic,amssymb, nobibnotes, aps, prd]{revtex4-1}

\usepackage[utf8]{inputenc}
\usepackage[english]{babel}
\usepackage[T1]{fontenc}
\usepackage{amsmath}
\usepackage{empheq}
\usepackage{amsfonts}
\usepackage{amssymb}
\usepackage{blkarray}
\usepackage{lmodern}
\usepackage{float} 
\usepackage{amsfonts}  
\usepackage{systeme}
\usepackage{enumitem}
\usepackage[squaren, Gray, cdot]{SIunits}
\usepackage{relsize,exscale}
\usepackage{mwe}
\usepackage{ragged2e}
\usepackage{breqn}
\usepackage{booktabs}

\DeclareMathOperator{\e}{e}
\DeclareMathOperator{\deriv}{d}

\begin{document}

\title{Linear analytical approach to dispersive, external and intrinsic dissipative couplings in
optomechanical systems}%

\author{J. Baraillon, B. Taurel, P. Labeye, and L. Duraffourg}%
\email[e-mail: ]{laurent.duraffourg@cea.fr}
\affiliation{Université Grenoble Alpes$,$ CEA$,$ LETI$,$ F38054 Grenoble}
\date{29 April 2020}%


\begin{abstract}
We present a theoretical study of optomechanical systems in which the mechanical resonator modulates both the resonant frequency (dispersive coupling) and the decay rates (dissipative coupling) of the optical cavity. We extend the generic dispersive framework to a more general case in which the dissipative coupling is split between its external and intrinsic contribution. We report a complete analysis of the influence of each kind of optical losses (intrinsic and external) on the three coupling mechanisms and highlight the interest of each external decay rate regime. The basic tools to experimentally identify the three couplings and their relative influence on the optical response are presented. We demonstrate the general expression of the optical spring effect and optomechanical damping. Comparison between experimental measurements in photonic crystal systems from the literature and our theoretical modal yields good agreement. 
\end{abstract}


\maketitle
\tableofcontents



\section{Introduction}

	Cavity optomechanics explores the mutual interaction of electromagnetic radiation and mechanical vibrations. In most optomechanical systems, the mechanical displacement modulates the resonant frequency of an optical cavity. This dispersive coupling gives rise to several phenomena, such as the optical spring effect and amplification or cooling of the mechanical motion which have been studied theoretically \cite{TH_aspelmeyer_2014,DISP_TH_marquardt_2007,DISP_TH_wilson_2007}, and have been achieved in various setups such as membrane inside a Fabry-Perot cavity (so-called ``membrane-in-the-middle'') \cite{DISP_jayich_2008}, suspended microdisk \cite{DISP_ding_2010} and phoxonic crystal system \cite{DISP_chan_2012}. Dissipative coupling, where the photon decay rate is modulated by mechanical vibrations, can also arise in optomechanical systems. This scheme was first proposed theoretically as an alternative to dispersive coupling in the so-called unresolved sideband limit \cite{TH_elste_2009}, and  also for its squeezing capability \cite{SQ_gu_2013,SQ_kronwald_2013,SQ_tagantsev_2018}. This coupling mechanism was implemented experimentally in diverse configurations such as Michelson-Sagnac interferometers \cite{MS_xuereb_2011,MS_tarabrin_2013,MS_sawadsky_2015,MS_nazmiev_2019}, whispering gallery mode resonators coupled to a nanomechanical beam waveguide \cite{WGM_li_2009,WGM_huang_2010,WGM_fu_2012,WGM_madugani_2015} , ring resonators coupled to a micromechanical resonator \cite{RR_huang_2018} and in photonic crystal (PhC) systems \cite{PC_wu_2014,PC_hryciw_2015}.

\begin{figure}[h]
        \centering
        \includegraphics[scale = 1.28]{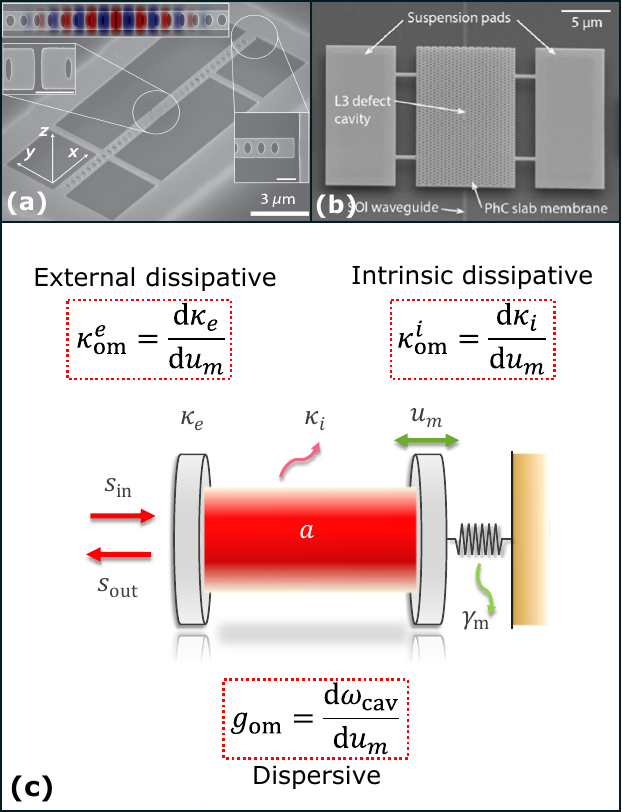}   
        \caption{\textbf{(a)} Scanning Electron Microscope (SEM) image of a PhC split-beam nanocavity from Wu \textit{et al.} \cite{PC_wu_2014} \textbf{(b)} SEM image of an integrated PhC mechanical resonator vertically stacked over a silicon on insulator waveguide from Tsvirkun \textit{et al.} \cite{PC_tsvirkun_2015} \textbf{(c)} Schematic illustration of the equivalent Fabry-Perot optical cavity with one movable mirror. The important quantities, namely the normalized field amplitudes ($a_{\textup{in}}$, $a_{\textup{out}}$ and $a$), the mechanical displacement ($u_m$), the decay rates ($\kappa$, $\kappa_e$ and $\kappa_i$), the optical resonance frequency $\omega_{cav}$, and the three first order optomechanical couplings ($g_{\textup{om}}$, $\kappa_{\textup{om}}^e$ and $\kappa_{\textup{om}}^i$), are introduced.} 
        \label{fig:fig1}
\end{figure}

More recently, it has been shown that dissipative coupling must be split into its external ($\kappa_{\textup{om}}^e$) and intrinsic ($\kappa_{\textup{om}}^i$) contribution, which have been studied theoretically \cite{TH_weiss_3couplings_2013} and experimentally in two different PhC systems \cite{PC_wu_2014,PC_hryciw_2015,PC_tsvirkun_2015}. Wu \textit{et al.}, in their PhC split-beam nanocavity (figure \ref{fig:fig1} (a)), measured dispersive coupling ($g_{\textup{om}}$) and $\kappa_{\textup{om}}^i$ in the $\unit{}{\giga\hertz\per\nano\meter}$ range and $\kappa_{\textup{om}}^e$ in the $\unit{}{\mega\hertz\per\nano\meter}$ range. While the external dissipative coupling is weaker, they observed that it has a larger impact in their external undercoupled regime (unlike the overcoupled regime considered by Elste \textit{et al.} in their theoretical study \cite{TH_elste_2009}), meaning that a change in the external decay rate has a large influence on the optical response. Tsvirkun \textit{et al.}, in their PhC slab suspended over an input waveguide (figure \ref{fig:fig1} (b)), measured $g_{\textup{om}} $, $\kappa_{\textup{om}}^e$ and $\kappa_{\textup{om}}^i$ in the $\unit{}{\giga\hertz\per\nano\meter}$ range. They observed various detuning behavior depending on the external decay rate, by changing the geometrical characteristics of the input waveguide. Additionaly, although Elste \textit{et al.} considered that dissipative coupling occurs solely through the modulation of the external optical losses (see erratum of \cite{TH_elste_2009}), several coupling configurations have been achieved experimentally by Tsvirkun \textit{et al.}, including a situation where intrinsic dissipative coupling is greater than the two other couplings.

	In this paper, we extend the theoretical framework based on the formalism used to describe purely dispersive optomechanical systems (see \cite{TH_aspelmeyer_2014, DISP_TH_marquardt_2007, DISP_TH_wilson_2007}). Our goal is to give a general description of the optomechanical interactions in presence of dispersive, external and intrinsic dissipative couplings. The tools introduced here can be used in various future applications where the distinction between the two dissipative couplings is necessary. We constantly refer to the PhC systems of Wu \textit{et al.} \cite{PC_wu_2014} and of Tsvirkun \textit{et al.} \cite{PC_tsvirkun_2015} to illustrate our calculations in concrete situations. In section \ref{sct2}, we develop the optical input-output relation and the mechanical dynamical equation in order to find the general coupled equations of motion. In section \ref{sct3}, we analyse the mean ouput optical response to understand the relative influence of each coupling under different external decay rate regimes, highlighting the interest of each configuration on the optical readout of the mechanical motion. It will help to easily estimate the optical sensitivity of future optomechanical systems and to experimentally identify the coupling strength of each contribution. In section \ref{sct4}, we derive the impact of each coupling on the intracavity field fluctuations around the mean field value and use the backaction force operator to determine the mechanical spectrum with the usual optomechanical effects (optically induced frequency shift and optomechanical damping) which completes the previous work of Weiss et al. \cite{TH_weiss_3couplings_2013}. The impact of optical detuning on these quantities are finally compared to the measurements of Tsvirkun et al \cite{PC_tsvirkun_2015}.

\section{Coupled equations of motion}
\label{sct2}

In this section we introduce the coupled equations of motion in presence of three optomechanical couplings. The generic theory developped here is only valid in the linear case \textit{i.e.} for small mechanical displacements (first order perturbations). We consider a generic optomechanical system, constituted by an optical and a mechanical resonators of respective resonant frequencies $\omega_c$ and $\omega_m$. The optical element is considered to be a one-port cavity in which the light can be coupled in or out by the same side. Typical one-port optical systems are a Fabry-Pérot with one partially reflective mirror and one perfectly reflective mirror (see figure \ref{fig:fig1} (c)), all-pass ring resonators \cite{RR_huang_2018}, and whispering gallery mode resonators coupled to an optical fiber \cite{WGM_madugani_2015}. Note that for multi-port optical systems, the model can still be valid by defining  different optical decay rates. The temporal dynamic of the intracavity complex field amplitude $a(t)$ is governed by the input-output relation \cite{TH_aspelmeyer_2014, TH_gardiner_2004}:

\begin{dmath}
\label{eqn:input_output_relations_intracavity}
\dot{a}(t) = -\big( \dfrac{\kappa(u_m)}{2} - i(\omega_L-\omega_c(u_m)) \big) \, a(t) + \sqrt{\kappa_{e}(u_m)}s_{\textup{in}}(t) 
\end{dmath}

\noindent where $\kappa_e$ and $\kappa_i$ are respectively the external and intrinsic photon decay rates, $\kappa = \kappa_e + \kappa_i$ is the  overall cavity decay rate, $\omega_L$ is the input laser angular frequency and $s_{\textup{in}}$ is the input laser flux. Note here that the field amplitude is normalised such that $|a|^2=n_{\textup{cav}}$ stands for the intracavity photon number. As in most optomechanical systems the optical frequency depends on the temporal mechanical amplitude $u_m(t)$. In our study we also consider that the decay rates depend on $u_m$. Therefore we introduce the dispersive coupling $g_{\textup{om}}$, external dissipative coupling $\kappa_{\textup{om}}^e$ and intrinsic dissipative coupling $\kappa_{\textup{om}}^i$ which correspond to a shift of respectively the cavity resonance frequency, the external decay rate and the intrinsic decay rate due to the mechanical oscillator motion, and are defined in the first order Taylor expansion as

\begin{dgroup*}[noalign]
\begin{dmath}
\label{eqn:couplings}
\omega_c(u_m) = \omega_{c0} + g_{\textup{om}} u_m(t) \, ,
\end{dmath}
\begin{dmath}
\kappa_e(u_m) = \kappa_{e0} + \kappa_{\textup{om}}^e u_m(t) \, ,
\end{dmath}
\begin{dmath}
\kappa_i(u_m) = \kappa_{i0} + \kappa_{\textup{om}}^i u_m(t) \, ,
\end{dmath}
\end{dgroup*}

\noindent with $\omega_{c0}$, $\kappa_{e0}$ and $\kappa_{i0}$ respectively the optical bare cavity (\textit{i.e.} without any optomechanical interaction) angular frequency, external and intrinsic decay rates. Note here that while quadratic coupling can be relevant in some situations, such as in the ``membrane-in-the-middle'' setup where a purely dispersive quadratic coupling could lead to quantum non-demolition measurement of the mechanical ground state \cite{DISP_jayich_2008}, it is beyond the scope of this study and we will focus only on the effects induced by the three kind of first order couplings.

The mechanical element is modeled as an harmonic oscillator with intrinsic damping $\gamma_m = \omega_m/Q_m$ (with $Q_m$ the mechanical quality factor) and effective mass $m_{\textup{eff}}$. The dynamical temporal behavior of the mechanical complex amplitude is governed by the equation \cite{TH_aspelmeyer_2014, MEM_hauer_2013}

\begin{dmath}
\label{eqn:mechanical_equation}
\begin{array}{l}
\ddot{u_m} (t) + \gamma_m\dot{u_m} (t) + \omega_m^2  u_m (t) = \dfrac{ F_L(t)}{m_{\textup{eff}}} + \dfrac{ F_{\textup{opt}}(t)}{m_{\textup{eff}}} \, ,
\end{array}
\end{dmath}

\noindent where $F_L(t)$ represents the thermal Langevin force arising from the thermal fluctuations and responsible for the brownian motion of the mechanical resonator and $F_{\textup{opt}}(t)$ represents the optical force induced by the intracavity field. 

In the following we expand and linearize the input-ouput relation for the intracavity field, in order to derive the coupled optomechanical equations of motion. Consider the steady-state value of field amplitude $\bar{a}$ and mechanical displacement $\bar{u}_m$ and respective temporal fluctuations $\delta a(t)$ and $\delta u_m(t)$ around these mean values such that $a(t) = (\, \bar{a}+\delta a(t)\, ) \e^{-i\omega_L t}$ and $u_m(t) = \bar{u}_m+\delta u_m(t)$. Second order terms such as $\delta a(t) \delta u_m(t)$ are neglected. We consider a continuous input flux such that $s_{\textup{in}} = \bar{s}_{\textup{in}}\e^{-i\omega_Lt}$ with $\bar{s}_{\textup{in}}$ constant (with external modulation of the input field additional terms may be added to the following equations). We define $\bar{\Delta} = \Delta_0 - g_{\textup{om}} \bar{u}_m$ the effective optical detuning (with $\Delta_0 = \omega_L-\omega_{c0}$), $\bar{\kappa}_{e0} = \kappa_{e0} + \kappa_{\textup{om}}^e \bar{u}_m$ and $\bar{\kappa}_{i0} = \kappa_{i0} + \kappa_{\textup{om}}^i \bar{u}_m$ the effective external and intrinsic photon decay rates and $\bar{\kappa} = \bar{\kappa}_{e0} + \bar{\kappa}_{i0}$ the effective overall photon decay rate. Note that non linear effect named static bistability, which has been studied elsewhere \cite{TH_aspelmeyer_2014}, arises because of the impact of the mean mechanical displacement on the optical parameters. Under the assumption of small mechanical displacement fluctuations, we can use the first order expansion $\sqrt{\kappa_e(u_m)} = \sqrt{\bar{\kappa}_{e0}}(1 + (\kappa_{\textup{om}}^e  / 2\bar{\kappa}_{e0}) \, \delta u_m(t) )$ to obtain equations \ref{eqn:mean_values_opt} to \ref{eqn:fluctuations_differential_equations_meca}. The mean intracavity field and displacement amplitudes are given by

\begin{dgroup*}[noalign]
\begin{dmath}
\label{eqn:mean_values_opt}
\bar{a} = \dfrac{\sqrt{\bar{\kappa}_{e0}}}{\bar{\kappa}/2 - i\bar{\Delta}} \, \bar{s}_{\textup{in}} \, ,
\end{dmath}
\begin{dmath}
\label{eqn:mean_values_meca}
\bar{u}_m = \dfrac{ \bar{F}_{\textup{opt}}}{m_{\textup{eff}}\omega_m^2} \, ,
\end{dmath}
\end{dgroup*}

\noindent where  $\bar{F}_{\textup{opt}}$ represents the mean optical force responsible for the change in the detuning and the decay rates. Therefore the field and displacement fluctuations obey the following differential equations:

\begin{dmath}
\label{eqn:fluctuations_differential_equations_opt}
	\dot{\delta a}(t) = \left(-\dfrac{\bar{\kappa}}{2}+ i\bar{\Delta} \right) \delta a(t) - \left[ \left( \dfrac{\kappa_{\textup{om}}^e + \kappa_{\textup{om}}^i}{2} + ig_{\textup{om}} \right) \bar{a} - \dfrac{\kappa_{\textup{om}}^e}{2\sqrt{\bar{\kappa}_{e0}}} \, \bar{s}_{\textup{in}} \right] \delta u_m(t)  + \sqrt{\bar{\kappa}_{e0}} \, \bar{s}_{\textup{in}} \, ,	
\end{dmath}

\begin{dmath}
\label{eqn:fluctuations_differential_equations_meca}
	\ddot{\delta u_m} (t) + \gamma_m\dot{\delta u_m} (t) + \omega_m^2 \delta u_m (t)  = \dfrac{\delta F_L(t)}{m_{\textup{eff}}}  +  \dfrac{\delta F_{\textup{opt}}(t)}{m_{\textup{eff}}} \, .
\end{dmath}

While this last set of equations is similar to the purely dispersive optomechanical coupled equation of motion \cite{TH_aspelmeyer_2014,DISP_TH_marquardt_2007,DISP_TH_wilson_2007}, we can observe that the dissipative couplings give rise to supplementary terms proportionnal to the mechanical displacement fluctuations. As observed in a previous study on dissipative optomechanical systems \cite{TH_weiss_2013}, and in contrast to the dispersive case which is a purely cavity assisted coupling proportional to the intracavity steady-state field $\bar{a}$, the coupling here is also proportionnal to the drive flux $s_{\textup{in}}$. Note that contrary to this study, we distinguish the intrinsic and external components, which allows us to trace this effect back to the external dissipative coupling influence.

In the following we study the influence of the three couplings on the mean optical response. Then the two coupled equations of motion are used as a starting point to derive the mechanical spectrum and the optomechanical effects in presence of dispersive, external and intrinsic dissipative couplings.
\medskip


\section{Output optical response}
\label{sct3}

\begin{figure*}
        \includegraphics[width=0.79\textwidth]{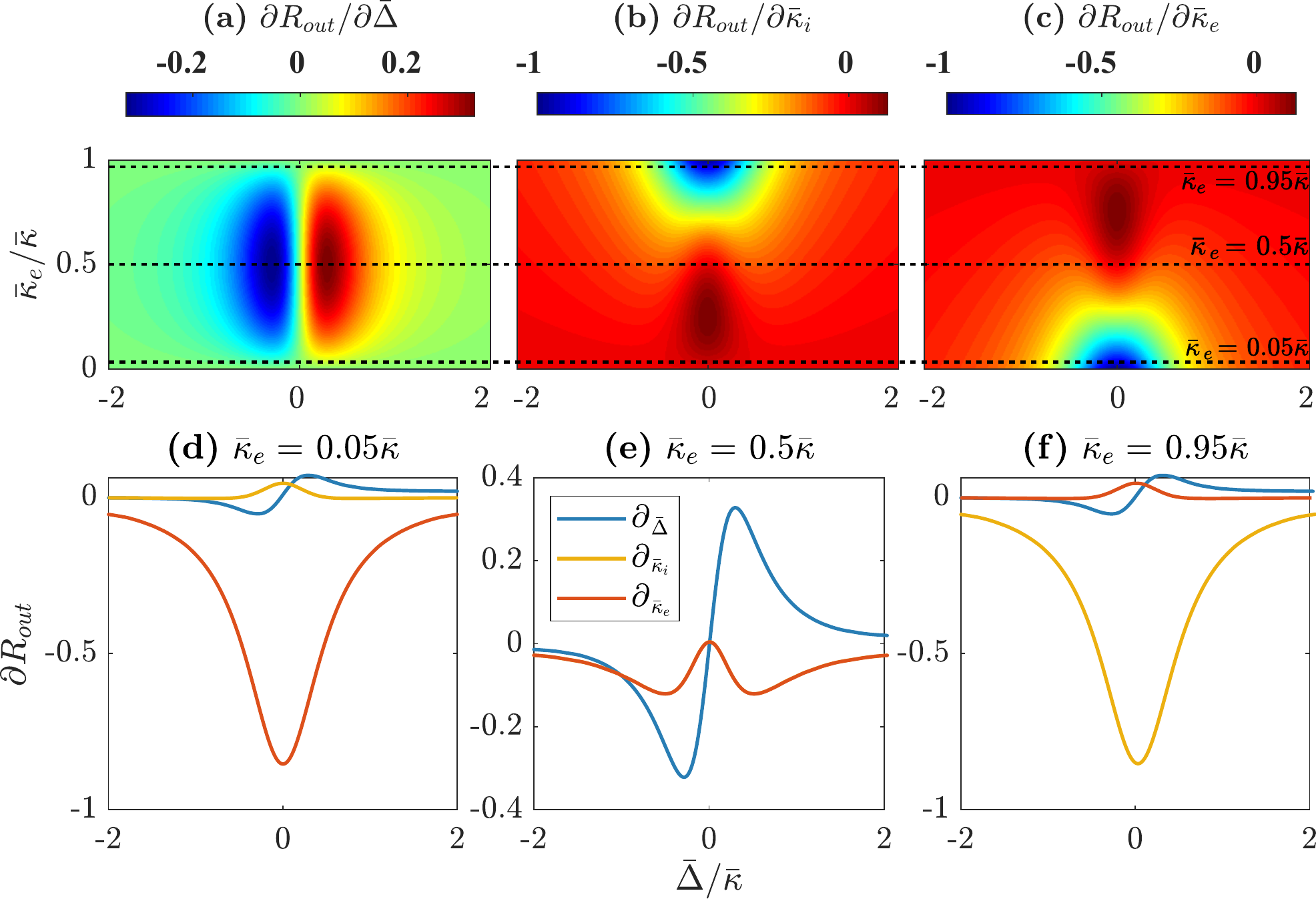} 
        \caption[ ]
        { Derivatives of the mean optical response amplitude as a function of the normalized detuning $\bar{\Delta}/\bar{\kappa}$. Influence of the normalized external cavity decay rate $\bar{\kappa}_{e0}/\bar{\kappa}$ on \textbf{(a)} $\partial R_{\textup{out}}/\partial \bar{\Delta}$, \textbf{(b)} $\partial R_{\textup{out}}/\partial \bar{\kappa}_{i0}$ and \textbf{(c)} $\partial R_{\textup{out}}/\partial \bar{\kappa}_{e0}$. Comparison between the three derivative amplitudes in \textbf{(d)} undercoupled regime, \textbf{(e)} critically coupled regime (green and blue curves are overlapping) and \textbf{(f)} overcoupled regime. For the six plots, the amplitudes are normalized with the maximum value between the three derivatives. The comparison made here is purely qualitative and is independant of the value of $\bar{\kappa}$.} 
        \label{fig:fig2}
\end{figure*}

In this section, we analyse the impact of each coupling scheme on the steady-state output optical response. As we considered a one port optical cavity, there is a single output photon flux $s_{\textup{out}}$. The input-output relation for this output flux is given by:

\begin{dmath}
\label{eqn:input_output_relations_output}
s_{\textup{out}}(t) = \sqrt{\kappa_{e}(u_m)} a(t) - \bar{s}_{\textup{in}} \, . 
\end{dmath}

We linearize this relation around a steady state value $\bar{s}_{\textup{out}}$. The steady state output optical flux is deducted from equation \ref{eqn:mean_values_opt}:

\begin{dmath}
\label{eqn:output_mean_value}
\bar{s}_{\textup{out}} = \dfrac{\bar{\kappa}_{e0} - \bar{\kappa}/2 + i\bar{\Delta}}{\bar{\kappa}/2 - i\bar{\Delta}} \, \bar{s}_{\textup{in}} \, .
\end{dmath}

The steady-state output response, defined as $R_{\textup{out}} = |\bar{s}_{\textup{out}} / \bar{s}_{\textup{in}}|^2$ is given by the following transfer function:

\begin{dmath}
\label{eqn:output_response}
R_{\textup{out}} = \dfrac{(\bar{\kappa}_{e0} - \bar{\kappa}/2)^2 + \bar{\Delta}^2}{(\bar{\kappa}/2)^2 + \bar{\Delta}^2} \, . 
\end{dmath}

$R_{\textup{out}}$ is similar to the typical reflection response of a one-port Fabry-Pérot cavity (with one partially and the other perfectly reflective mirrors). We now are going to determine the mechanically induced optical oscillations by using the method described in appendix of \cite{PC_wu_2014}:

\begin{dmath}
\label{eqn:output_response_oscillations}
\dfrac{\deriv R_{\textup{out}}}{\deriv u_m} = g_{\textup{om}}  \dfrac{\partial R_{\textup{out}}}{\partial \bar{\Delta}} + \kappa_{\textup{om}}^{e}  \dfrac{\partial R_{\textup{out}}}{\partial \bar{\kappa}_{e}} + \kappa_{\textup{om}}^{i}  \dfrac{\partial R_{\textup{out}}}{\partial \bar{\kappa}_{i}} \, ,
\end{dmath}

\noindent where the derivatives of $R_{\textup{out}}$ are given by: \smallskip

\begin{dgroup*}
\begin{dmath}
\label{eqn:output_response_Rout_derivative_disp}
\dfrac{\partial R_{\textup{out}}}{\partial \bar{\Delta}} = \dfrac{2\bar{\Delta}(1-R_{\textup{out}})}{(\bar{\kappa}/2)^2 + \bar{\Delta}^2} \, ,
\end{dmath}
\begin{dmath}
\label{eqn:output_response_Rout_derivative_diss_e}
\dfrac{\partial R_{\textup{out}}}{\partial \bar{\kappa}_{e}} = \dfrac{(\bar{\kappa}/2-\bar{\kappa}_{e0})-(\bar{\kappa}/2)R_{\textup{out}}}{(\bar{\kappa}/2)^2 + \bar{\Delta}^2} \, ,
\end{dmath}
\begin{dmath}
\label{eqn:output_response_Rout_derivative_diss_i}
\dfrac{\partial R_{\textup{out}}}{\partial \bar{\kappa}_{i}} = \dfrac{(\bar{\kappa}_{e0} - \bar{\kappa}/2)-(\bar{\kappa}/2)R_{\textup{out}}}{(\bar{\kappa}/2)^2 + \bar{\Delta}^2} \, .
\end{dmath}
\end{dgroup*}

In the following, a qualitive analysis of the detuning dependency of these three quantities for different external decay rate configuration (\textit{i.e.} value of $\bar{\kappa}_{e0}$) is given. We identify the external decay rate regimes where the optical readouts of each optomechanical interaction, namely dispersive, intrinsic and external, are enhanced. A more quantitative study is then given, with different configurations of optomechanical coupling values and their impact on the amplitude of the mechanically induced mean optical power oscillations.

\subsection{External decay rate regimes}

The derivatives of the mean optical response as a function of the normalized detuning $\bar{\Delta}/\bar{\kappa}$ are shown on figure \ref{fig:fig2}. The influence of the external cavity decay rate (\textit{i.e.} the external coupling coefficient) is shown on figure \ref{fig:fig2} \textbf{(a)}, \textbf{(b)} and \textbf{(c)}. These three figures highlight the most important discrepancy between the dispersive and dissipative derivatives : the dispersive behavior is characterized by two off resonant sidebands whereas the dissipative (both intrinsic and external) behavior is characterized by one resonant maximum. We can thus identify three particular external decay rate regimes:

\begin{itemize}
\item undercoupled regime: $\bar{\kappa}_{e0} \ll \bar{\kappa} \approx \bar{\kappa}_{i0}$ (Fig. \ref{fig:fig2}  \textbf{(d)}),
\item critically coupled regime: $\bar{\kappa}_{e0} = \bar{\kappa}/2 = \bar{\kappa}_{i0}$ (Fig. \ref{fig:fig2}  \textbf{(e)}),
\item overcoupled regime: $\bar{\kappa}_{e0} \approx \bar{\kappa} \gg \bar{\kappa}_{i0}$ (Fig. \ref{fig:fig2}  \textbf{(f)}),
\end{itemize}

\noindent and compare the detuning dependency of the three derivatives in these regimes in figure \ref{fig:fig2} \textbf{(d)}, \textbf{(e)} and \textbf{(f)}.

In the critically coupled regime (see figure \ref{fig:fig2} \textbf{(e)}), the dispersive variation (blue curve) of the mean optical response is the predominant mechanism with the highest impact at off resonance detunings ($\bar{\Delta}=\pm\bar{\kappa}/2$). In the undercoupled regime (see figure \ref{fig:fig2} \textbf{(d)}), intrinsic losses (\textit{i.e.} $\bar{\kappa}_{i0}$) are higher and the external dissipative variation (red curve) is the predominant mechanism. Therefore on resonance ($\bar{\Delta}=0$), a small variation of the intrinsic cavity decay rate $\bar{\kappa}_{i}$ will not have a strong influence on the optical output response $R_{\textup{out}}$, but a small variation of external cavity decay rate $\bar{\kappa}_{e}$ can lead an important impact on $R_{\textup{out}}$. In \cite{PC_wu_2014}, the system is in this undercoupled regime ($\bar{\kappa}=\unit{31}{\giga\hertz}$ and $\bar{\kappa}_{e0} = \unit{1}{\giga\hertz}$ \textit{i.e.} $\bar{\kappa}_{e0} = 0.03 \bar{\kappa}$), and they indeed observed a strong dependence of the mean optical response on the external decay rate. In the overcoupled regime, the detuning dependency of the intrinsic (resp. external) dissipative variations are identical to the detuning dependency of the external (resp. intrinsic) dissipative variations in the undercoupled regime, due to the mathematical symmetry between equations \ref{eqn:output_response_Rout_derivative_diss_e} and \ref{eqn:output_response_Rout_derivative_diss_i}. Note that, for each of the six figures, the derivatives have been normalized with the maximum value between $\partial R_{\textup{out}}/\partial \bar{\Delta}$, $\partial R_{\textup{out}}/\partial \bar{\kappa}_{e}$ and $\partial R_{\textup{out}}/\partial \bar{\kappa}_{i}$, as well as we normalized the detuning $\bar{\Delta}$ and external decay rate $\bar{\kappa}_{e0}$ with the overall decay rate $\bar{\kappa}$, such that the previous analysis is general and independant from the quality of the optical cavity \textit{i.e.} independant of $\bar{\kappa}$.

\begin{figure}
        \centering
        \includegraphics[scale = 0.63]{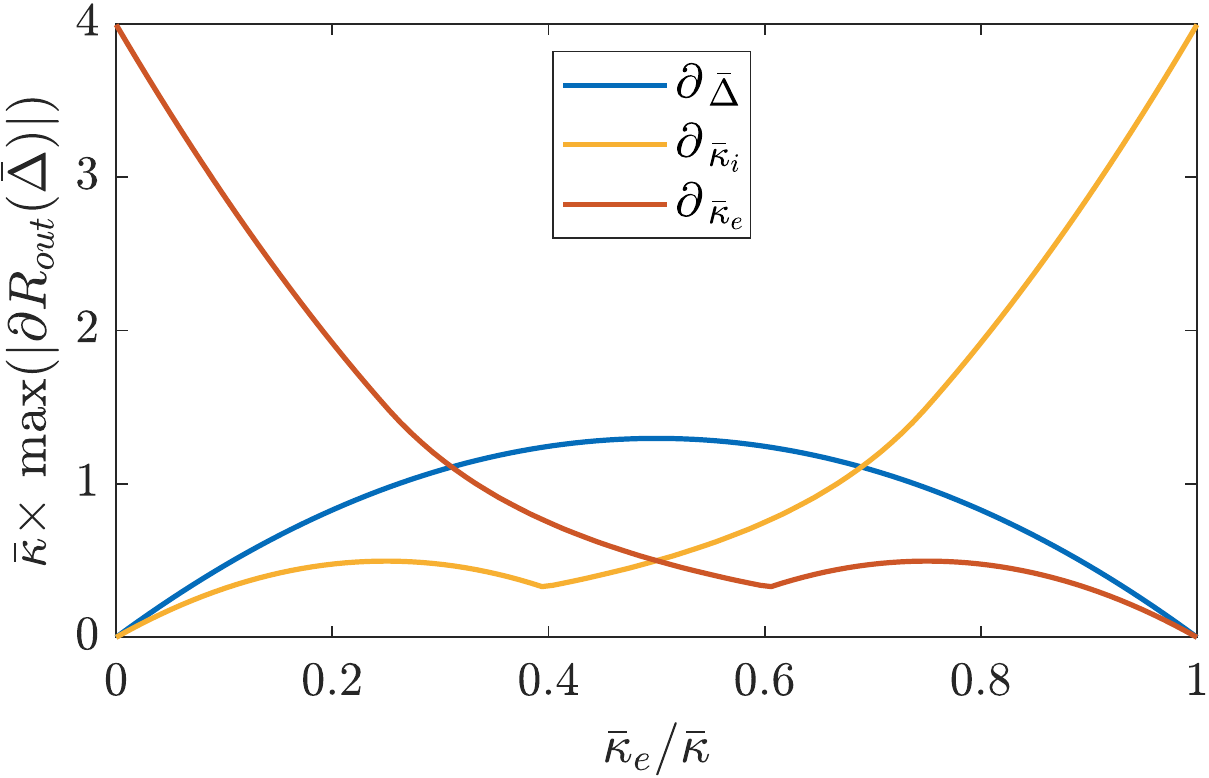}   
        \caption{Absolute maximum of the derivatives of the mean optical response as a function of the normalized external cavity decay rate $\bar{\kappa_e}/\bar{\kappa}$. Y axis is normalized such that y values are independant of $\bar{\kappa}$} 
        \label{fig:fig3}
\end{figure}

The dispersive derivative of the mean optical response ($\partial R_{\textup{out}} / \partial \bar{\Delta}$) is always an off-resonance effect associated with two sidebands, depending on whether we are interested on cooling or amplification of mechanical oscillations. Maximum of the dissipative derivatives ($\partial R_{\textup{out}} / \partial \bar{\kappa}_e$ and $\partial R_{\textup{out}} / \partial \bar{\kappa}_i$) are always on resonance. However, depending on the external coupling regime, two off resonance local maxima appear (close to the critically coupled regime, see red and yellow curve in figure \ref{fig:fig2} (e)), giving rise to new working sidebands. Figure \ref{fig:fig3} represents the maximum amplitude of the derivatives of $R_{\textup{out}}$ as a function of the external cavity decay rate. These maximum amplitudes are given by

\begin{dgroup*}
\begin{dmath}
\label{eqn:output_response_derivative_Rout_max_delta}
\left. \dfrac{\partial R_{\textup{out}}}{\partial \bar{\Delta}} \right|_{\textup{max}} =  \left|\dfrac{2\big(  1-R_{\textup{out}}(\bar{\Delta}=\pm\bar{\Delta}_{max}) \big)}{\bar{\kappa}}  \right| \overset{\bar{\kappa}_{e0} = \bar{\kappa}/2}{=}  \quad \dfrac{3\sqrt{3}}{4\bar{\kappa}} \, ,
\end{dmath}
\begin{dmath}
\label{eqn:output_response_derivative_Rout_max_ext}
\left. \dfrac{\partial R_{\textup{out}}}{\partial \bar{\kappa}_{e}} \right|_{\textup{max}} = \left| \dfrac{(\bar{\kappa}/2-\bar{\kappa}_{e0})-(\bar{\kappa}/2)R_{\textup{out}}(\bar{\Delta} = 0)}{(\bar{\kappa}/2)^2 } \right| \overset{\bar{\kappa}_{e0} \ll \bar{\kappa}}{=}  \quad \dfrac{4}{\bar{\kappa}} \, ,
\end{dmath}
\begin{dmath}
\label{eqn:output_response_derivative_Rout_max_int}
\left. \dfrac{\partial R_{\textup{out}}}{\partial \bar{\kappa}_{i}} \right|_{\textup{max}} = \left| \dfrac{(\bar{\kappa}_{e0}-\bar{\kappa}/2)-(\bar{\kappa}/2)R_{\textup{out}}(\bar{\Delta} = 0)}{(\bar{\kappa}/2)^2 } \right| \overset{\bar{\kappa}_{e0} = \bar{\kappa}}{=}  \quad \dfrac{4}{\bar{\kappa}} \, ,
\end{dmath}
\end{dgroup*}

\noindent where $\bar{\Delta}_{\textup{max}}=\bar{\kappa}/\sqrt{12}$ is the detuning of maximum slope of the mean optical response, which is determined by making $\partial^2 R_{\textup{out}}/\partial \bar{\Delta}^2$ equal to zero. The maximum amplitude of the derivatives of the mean optical response is therefore inversely proportional to the overall cavity decay rate, which highlights the interest of working in the resolved sideband regime ($\bar{\kappa} \ll \omega_m$) to increase optical sensibility towards mechanical displacement. We see that no matter the optical cavity decay rate $\bar{\kappa}$, the dissipative variations (both external and intrinsic) of the output response in the corresponding external decay rate regimes (respectively: undercoupled and overcoupled) are always more than 3 times higher ($16/3\sqrt{3} \approx 3.1$) than the dispersive one in the critically coupled regime, emphasizing the interest of dissipative coupling optical readout of the mechanical motion (for classical applications), in the corresponding external coupling regime.

\subsection{Mean optical power oscillations}

We quantitatively study the dependence of the  mean optical power oscillations of our generic system on detuning and external cavity decay rate for different optomechanical coupling configurations. The ouput power is defined as $P_{\textup{out}} = R_{\textup{out}}P_{\textup{in}}$ with $P_{\textup{in}}$ the input power. Based on equation \ref{eqn:output_response_oscillations}, we can determine the mechanically induced mean optical power oscillations in $\unit{}{\watt\per\meter}$: $\deriv P_{\textup{out}}/\deriv u_m = P_{\textup{in}}\deriv R_{\textup{out}}/\deriv u_m$. Note that we actually use an input power of $\unit{1}{\milli\watt}$ and thus the $\unit{}{\micro\watt\per\pico\meter}$ for convenience.

\begin{figure}
		\centering
        \includegraphics[width=0.35\textwidth]{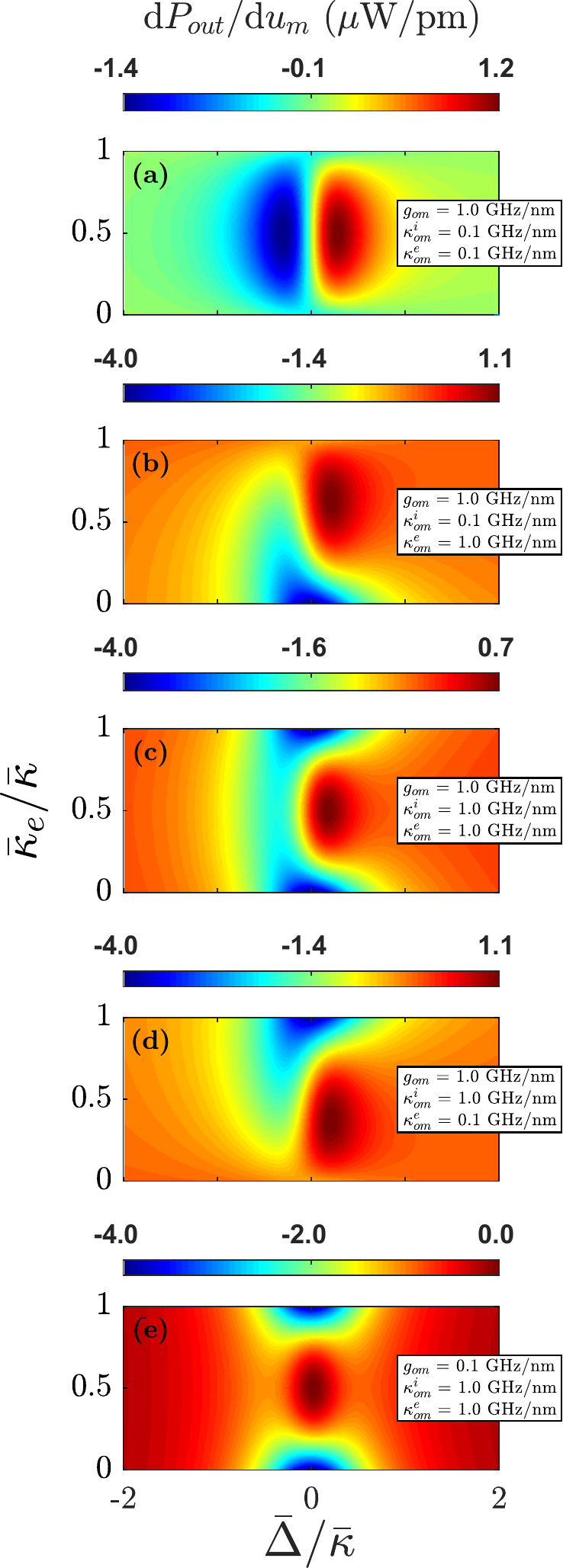}   
        \caption[]
        {Mechanically induced mean optical power oscillations as a function of the normalized detuning $\bar{\Delta}/\bar{\kappa}$ and the normalized external cavity decay rate $\bar{\kappa}_e/\bar{\kappa}$ for different optomechanical coupling configurations, in the unresolved sideband limit $\bar{\kappa} = 10^3\omega_m$ with $\omega_m=\unit{1}{\mega\hertz}$.} 
        \label{fig:fig4}
\end{figure}

As described in \cite{PC_wu_2014}, fiting the detuning dependency of the optical power on mechanical resonance (\textit{i.e.} the optical spectrum maximum amplitude) with equations \ref{eqn:output_response_oscillations}, \ref{eqn:output_response_Rout_derivative_disp}, \ref{eqn:output_response_Rout_derivative_diss_e} and \ref{eqn:output_response_Rout_derivative_diss_i}, allows to estimate the relative contribution (\textit{i.e.} the coupling strength $g_{\textup{om}}$, $\kappa_{\textup{om}}^i$ and $\kappa_{\textup{om}}^i$) of each optomechanical coupling process in an optomechanical experiment. This method uses the general expression of the output optical response and can thus be applied to a large variety of systems. Figure \ref{fig:fig4} represents the mechanically induced mean optical power oscillations $\deriv P_{\textup{out}}/\deriv u_m$ as a function of the optical detuning and the external cavity decay rate for five optomechanical coupling configurations (the coupling values are given on the figure).

\begin{table}[H]
\centering
\begin{tabular}{cccc}
\toprule
& \textbf{Wu \textit{et al.}} \cite{PC_wu_2014} && \textbf{Tsvirkun \textit{et al.}} \cite{PC_tsvirkun_2015} \\
\hline
$g_{\textup{om}} \unit{}{(\giga\hertz\per\nano\meter)}$ & $1,1-1,8$ && $0,62-1,52$ \\
$\kappa_{\textup{om}}^e \unit{}{(\giga\hertz\per\nano\meter)}$ & $0,002-0,003$ && $0,13-0,33$ \\
$\kappa_{\textup{om}}^i \unit{}{(\giga\hertz\per\nano\meter)}$ & $0,3-0,5$ &&  $0,01 - 5,64$ \\
$\bar{\kappa}/\omega_m$ & $10^3$ && $10^{3-4}$ \\
\bottomrule
\end{tabular}
\caption{Optomechanical coupling absolute values and sideband factor measured by Wu \textit{et al.} \cite{PC_wu_2014} and Tsvirkun \textit{et al.} \cite{PC_tsvirkun_2015}.}
\label{tab:Wu_Tsvirkun_couplings}
\end{table}

Table \ref{tab:Wu_Tsvirkun_couplings} summarizes the typical absolute coupling values and corresponding sideband parameter $\bar{\kappa}/\omega_m$ measured by Wu et. al in one of their PhC split-beam nanocavity \cite{PC_wu_2014,PC_hryciw_2015} and by Tsvirkun \textit{et al.} in their PhC slab suspended over an input waveguide \cite{PC_tsvirkun_2015, PC_tsvirkun_thesis_2015}. The best optomechanical systems, designed for dispersive optomechanical cooling, achieved $\bar{\kappa} = 0.02 \, \omega_m$ \cite{TH_aspelmeyer_2014}. However, based on Wu \textit{et al.} and Tsvirkun \textit{et al.} measurements, we observe that the sideband factor $\bar{\kappa}/\omega_m$ has no impact on the achievable dispersive and dissipative coupling values. The amplitude of mechanically induced mean optical power oscillations is therefore not influenced by this parameter, but only by the  value of $\bar{\kappa}$ itself \textit{i.e.} the quality of the optical cavity. It is also independent on the mechanical resonator properties (resonant frequency $\omega_m$, quality factor $Q_m$ and effective mass $m_{\textup{eff}}$). The sideband factor influences only the optically induced effects on the mechanical properties (see last section).

We choose $\bar{\kappa} = 10^3 \,\omega_m = \unit{1}{\giga\hertz}$ (close to one of the devices of Tsvirkun \textit{et al.} \cite{PC_tsvirkun_2015, PC_tsvirkun_thesis_2015}). Figure \ref{fig:fig4} shows that in this case the maximum absolute power oscillations are around $\unit{4}{\micro\watt\per\pico\meter}$, for $g_{\textup{om}}$, $\kappa_{\textup{om}}^e$ and $\kappa_{\textup{om}}^i$ at $\unit{1}{\giga\hertz\per\nano\meter}$, and that this value increases linearly when decreasing $\bar{\kappa}$ ($\unit{0.4}{\milli\watt\per\pico\meter}$ for $\bar{\kappa} = 10 \, \omega_m = \unit{10}{\mega\hertz}$). In absence of optomechanical effects, we can easily calculate the corresponding thermal optical spectrum. For instance, consider a mechanical resonator with $m_{\textup{eff}}= \unit{117.2}{\pico\gram}$ and $Q_m = 2000$ (from Tsvirkun \textit{et al.}, see table \ref{tab:parameters}). The associated resonant brownian motion at room temperature ($T = \unit{294}{\kelvin}$) is given by $\sqrt{S_{\textup{th}}} = \sqrt{4k_BTQ_m/m_{\textup{eff}}\omega_m^3}$ \cite{MEM_hauer_2013} where $k_B=\unit{1.38\times10^{-23}}{\meter^2\kilo\gram\per\second^2\kelvin}$ is the Boltzmann constant. Thus we have $\sqrt{S_{\textup{th}}}\sim\unit{0.16}{\pico\meter\per\sqrt{\hertz}}$, and the associated maximum optical power spectral density (PSD) is $\sqrt{S_{\textup{opt,th}}} \sim \unit{0.6}{\micro\watt\per\sqrt{\hertz}}$ for $\bar{\kappa} = \unit{1}{\giga\hertz}$ (and $\sqrt{S_{\textup{opt,th}}} \sim \unit{60}{\micro\watt\per\sqrt{\hertz}}$ for $\bar{\kappa} = \unit{10}{\mega\hertz}$). As a remark, we point out that the previous calculation is based on values from Tsvirkun \textit{et al.}, which altought typical, may vary according to the relative strenght of the optomechanical couplings. Nevertheless, the calculation remains relevant because it gives an order of magnitude of the quantity effectively measured.

Figure \ref{fig:fig4} (a) illustrates a mostly dispersive case where the strongest mean optical power oscillations arise at critical coupling and off resonant detuning. We consider relatively low dissipative couplings in comparison to $g_{\textup{om}}$, which explains the low power oscillations in the undercoupled and overcoupled regime. However, we observe an asymmetry in the order of magnitude of the dispersive sidebands due to the non-zero dissipative couplings. Indeed comparison with figure \ref{fig:fig2} reveals that the dissipative mechanisms induce, on resonance, a negative (or positive, depending of the sign of the couplings) amplification of the power oscillations which are added to or subtracted from the dispersive sidebands. This effect is responsible for the strong asymmetry of the five plots. Figure \ref{fig:fig4} (b) and (d) show the influence of a stronger respectively external and intrinsic dissipative coupling, with the same dispersive coupling value. We observe an amplification of the asymmetry in the critically coupled regime. The highest power oscillations in these two cases are achievable on resonance in the undercoupled (for higher external dissipative coupling) or overcoupled (for higher intrinsic dissipative coupling) regimes. Figure \ref{fig:fig4} (c) illustrates the case where the three optomechanical couplings are at the same level (Tsvirkun \textit{et al.} were close to this situation in one of their devices, see Figure 3 (d) in \cite{PC_tsvirkun_2015}). Finally, figure \ref{fig:fig4} (e) shows the dissipative case in which dispersive optomechanical coupling is 
negligible in comparison to the two other couplings, extinguishing almost completely the dispersive detuning sidebands in the critically coupled regime. In the last two situations, working in undercoupled or overcoupled regime will induce the same strong optical power oscillations on optical resonance. The sign of each coupling can also induce other discrepancies, but the behavior does not drastically change. The absolute mean optical power oscillations remains in the same order of magnitude, but maximum absolute values can arise at different detuning in the critically coupled regime. 

The method described in this section can be used to identify the strenght of each coupling in future optomechanical experiments, as Wu \textit{et al.} and Tsvirkun \textit{et al.} did in their own system. It can also be used in more practical applications, such as sensors, as a way to estimate the sensibility by comparing the mechanically induced optical power oscillations of interest with the noises induced by every components (laser source and electrical components) of the optomechanical experiment.

			 
\section{Mechanical spectrum}
\label{sct4}

We now study the dynamical mechanical response. We first use the input-output relation (see equation \ref{eqn:fluctuations_differential_equations_opt}) in order to determine the Fourier transform of the intracavity field fluctuations. After calculating the optical force in presence of dispersive and dissipative couplings, the mechanical response is determined, from which the general expression of the optically induced mechanical frequency shift and optomechanical damping are obtained. Finally, the theoretical optical spectrum in a concrete case is compared with previous measurements from Tsvirkun \textit{et al.} \cite{PC_tsvirkun_2015}.

From now on, we work in the Fourier space and choose the convention $a(\omega) = \int_{-\infty}^{+\infty} a(t) \e^{-i\omega t}$. The Fourier transform of equation \ref{eqn:fluctuations_differential_equations_opt} allows us to write the fluctuations of the intracavity field as

\begin{dmath}
\delta a(\omega) = \delta a_{\textup{disp}}(\omega) + \delta a_{\textup{diss}}^e(\omega) + \delta a_{\textup{diss}}^i(\omega) \, , \newline
\end{dmath}

\noindent with $ \delta a_{\textup{disp}}(\omega)$, $\delta a_{\textup{diss}}^e(\omega)$ and $\delta a_{\textup{diss}}^i(\omega)$ the fluctuations induced by respectively the dispersive, the external dissipative and intrinsic dissipative coupling given by

\begin{dgroup*}[noalign]
\begin{dmath}
\label{eqn:intracavity_fluctuations_disp}
	\delta a_{\textup{disp}}(\omega) = - ig_{\textup{\textup{om}}} \,  \chi_{\textup{cav}}^{\textup{eff}}(\omega) \, \bar{a} \, \delta u_m(\omega) \, ,
\end{dmath}
\begin{dmath}
\label{eqn:intracavity_fluctuations_diss_e}
	\delta a_{\textup{diss}}^e(\omega) = \dfrac{\kappa_{\textup{om}}^e( \bar{\kappa}/2-\bar{\kappa}_{e0} -i\bar{\Delta} )  }{2\bar{\kappa}_{e0}} \chi_{\textup{cav}}^{\textup{eff}}(\omega)    \bar{a}   \delta u_m(\omega)
\end{dmath}
\begin{dmath}
\label{eqn:intracavity_fluctuations_diss_i}
	\delta a_{\textup{diss}}^i(\omega) = - \dfrac{\kappa_{\textup{om}}^i}{2} \, \chi_{\textup{cav}}^{\textup{eff}}(\omega) \, \bar{a} \, \delta u_m(\omega) \, .
\end{dmath}
\end{dgroup*}

Here we recognize in each term the effective cavity response in presence of optomechanical interaction $\chi_{\textup{cav}}^{\textup{eff}}(\omega) = [\bar{\kappa}/2-i(\bar{\Delta}+\omega)]^{-1}$ which is due to the filtering role of the resonant optical cavity \cite{TH_aspelmeyer_2014}. The three couplings lead to an optical force $F_{\textup{opt}}(t)$ whose fluctuations can be described by the backaction force operator \cite{TH_elste_2009, WGM_li_2009}, which in the case of a constant input flux (\textit{i.e.} $\bar{s}_{\textup{in}}(t) = \bar{s}_{\textup{in}}$) writes:

\begin{dmath}
\label{eqn:optical_force}
\delta F_{\textup{opt}}(t) = -\hbar g_{\textup{\textup{om}}} \big( \bar{a}^{\ast}\delta a(t) + \bar{a}\delta a^{\ast}(t) \big) - i\hbar \dfrac{\kappa_{\textup{om}}^e + \kappa_{\textup{om}}^i}{2} \, \dfrac{s_{\textup{in}}}{\sqrt{\bar{\kappa}_{e0}}} \, \big( \delta a^{\ast}(t) - \delta a(t) \big) \, .
\end{dmath}

The first term corresponds to the dispersive optical force and is linked to the intracavity photon energy, which varies with mechanical displacement. The second term is the dissipative optical force analogous to a viscous force and linked to the photons leaking out of the cavity via external or intrinsic dissipation mechanisms. This general expression leads to a linear relation between Fourier transforms of the optical force and displacement fluctuations. The effective mechanical susceptibility $\chi_m^{\textup{eff}}$ defined by $\delta u_m(\omega, \bar{\Delta}) = \chi_m^{\textup{eff}}(\omega, \bar{\Delta}) F_{L}(\omega)$ is thus determined:

\begin{dgroup*}[noalign]
\begin{dmath}
\label{eqn:effective_susceptibility}
	\chi_m^{\textup{eff}}(\omega, \bar{\Delta}) = \dfrac{1}{m_{\textup{eff}} \big( \omega_m^2 - \omega^2 - i\omega\gamma_m \big) + \Sigma(\omega, \bar{\Delta}) } 
\end{dmath}
\begin{dmath}
\textup{where} \qquad \Sigma(\omega, \bar{\Delta})  =  -\dfrac{\delta F_{\textup{opt}}}{\delta u_m}
\end{dmath}
\end{dgroup*}

\noindent is the optomechanical self-energy \cite{TH_aspelmeyer_2014}. We can introduce the optically induced mechanical frequency shift $\delta \omega_m$ and optomechanical damping $\gamma_{\textup{opt}}$ with the relation $\Sigma = m_{\textup{eff}} ( 2\omega\delta \omega_{m} - i\omega \gamma_{\textup{opt}} )$. These two quantities are then expressed as

\begin{dgroup*}[noalign]
\begin{dmath}
	\delta \omega_{m}(\omega, \bar{\Delta})  = -\dfrac{1}{2\omega m_{\textup{eff}}} \, \mathbf{Re} \left[ \dfrac{\delta F_{\textup{opt}}}{\delta u_m} \right] \, ,
\end{dmath}
\begin{dmath}
	\gamma_{\textup{opt}}(\omega, \bar{\Delta}) =   \dfrac{1}{\omega m_{\textup{eff}}} \, \mathbf{Im} \left[ \dfrac{\delta F_{\textup{opt}}}{\delta u_m} \right] \, ,
\end{dmath}
\end{dgroup*}

\noindent where $\mathbf{Re}$ and $\mathbf{Im}$ respectively stand for real and imaginary part. As the optical force is composed of three forces induced by each coupling, the optically induced effects are composed of three terms proportional to $g_{\textup{\textup{om}}}^2$, $(\kappa_{\textup{om}}^e)^2$ and $(\kappa_{\textup{om}}^i)^2$ corresponding respectively to the purely dispersive,  external dissipative or intrinsic dissipative situation (see appendix). However as the intracavity field fluctuations depend on the three couplings at the same time (see equations (\ref{eqn:intracavity_fluctuations_disp}, \ref{eqn:intracavity_fluctuations_diss_e} and \ref{eqn:intracavity_fluctuations_diss_i})), there are also ``interference'' between them which lead to intertwined terms proportional to $g_{\textup{\textup{om}}}\kappa_{\textup{om}}^e$, $g_{\textup{\textup{om}}}\kappa_{\textup{om}}^i$ and $\kappa_{\textup{om}}^e\kappa_{\textup{om}}^i$ (see appendix).

The sum of the contributions of the purely dispersive, external and intrinsic dissipative, and the crossing terms lead to the overall optical spring effect and optomechanical damping in the presence of the three couplings:

\begin{dgroup*}[noalign]
\begin{dmath}
	\delta\omega_m   =  \delta \omega_m^{disp} + \delta \omega_m^{diss, e}(\omega) + \delta \omega_m ^{diss, i}  +  \delta \omega_m^{disp,diss,e} + \delta\omega_m^{disp,diss,i} + \delta \omega_m^{diss,e,diss,i}
	\label{eqn:optical_spring}
\end{dmath}
\end{dgroup*}

\begin{dgroup*}[noalign]
\begin{dmath}
	\gamma_{\textup{opt}}  =  \gamma_{\textup{opt}}^{disp} + \gamma_{\textup{opt}}^{diss, e} + \gamma_{\textup{opt}}^{diss, i} +  \gamma_{\textup{opt}}^{disp,diss,e} + \gamma_{\textup{opt}}^{disp,diss,i} + \gamma_{\textup{opt}}^{diss,e,diss,i}
	\label{eqn:optical_damping}
\end{dmath}
\end{dgroup*}

In practical experiments, we have access to the optical power spectral density (PSD) $S_{\textup{opt}}(\omega, \bar{\Delta})$ in $\unit{}{\watt^2\per\hertz}$, related to the mechanical PSD $S_m(\omega, \bar{\Delta})$ in $\unit{}{\meter^2\per\hertz}$ by 

\begin{dmath}
	S_{\textup{opt}}(\omega, \bar{\Delta}) = P_{\textup{in}}^2 \left|\dfrac{\deriv R_{out}}{\deriv u_m}\right|^2 S_m(\omega, \bar{\Delta}) \, ,
\end{dmath}

\noindent with $\deriv R_{out}/\deriv u_m$ given by equation \ref{eqn:output_response_oscillations}. The mechanical PSD is given by the fluctuation dissipation theorem \cite{TH_aspelmeyer_2014, MEM_hauer_2013}:

\begin{dmath}
	S_m(\omega, \bar{\Delta}) =  \dfrac{4k_BT\omega_m}{Q_m}|\chi_m^{\textup{eff}}(\omega)|^2
\end{dmath}

\begin{table}
\centering
\begin{tabular}{cll}
\toprule
\multicolumn{3}{c}{\textbf{Mechanical properties}} \\
\midrule
$\omega_m$ & \multicolumn{2}{c}{$\unit{2.22}{\mega\hertz}$} \\
$Q_m$ & \multicolumn{2}{c}{2000} \\
$m_{\textup{eff}}$ & \multicolumn{2}{c}{$\unit{117.2}{\pico\gram}$} \\
\bottomrule
\multicolumn{3}{c}{\textbf{Optical properties}} \\
\midrule
$P_{\textup{in}}$ & \multicolumn{2}{c}{$\unit{6.8}{\milli\watt}$} \\
$\lambda_{\textup{cav}}$ & \multicolumn{2}{c}{$\unit{1563.42}{\nano\meter}$} \\ 
\hline
$\bar{\kappa}$ & $\unit{0.09}{\nano\meter}$ \textit{i.e.} $\unit{11}{\giga\hertz}$ & (Fig. \ref{fig:fig5}) \\
& $\unit{0.18}{\nano\meter}$ \textit{i.e.} $\unit{22}{\giga\hertz}$ & (Fig. \ref{fig:fig6} \textbf{(a)}, \textbf{(b)}) \\
& $\unit{1.55}{\nano\meter}$ \textit{i.e.} $\unit{190}{\giga\hertz}$ & (Fig. \ref{fig:fig6} \textbf{(c)}, \textbf{(d)}) \\
& $\unit{0.72}{\nano\meter}$ \textit{i.e.} $\unit{88}{\giga\hertz}$ & (Fig. \ref{fig:fig6} \textbf{(e)}, \textbf{(f)}) \\ 
\hline
$\bar{\kappa}_e$ & $0.16\,\bar{\kappa}$ & (Fig. \ref{fig:fig5}) \\
& $0.08\,\bar{\kappa}$ & (Fig. \ref{fig:fig6} \textbf{(a)}, \textbf{(b)}) \\
& $0.21\,\bar{\kappa}$ & (Fig. \ref{fig:fig6} \textbf{(c)}, \textbf{(d)}) \\
& $0.10\,\bar{\kappa}$ & (Fig. \ref{fig:fig6} \textbf{(e)}, \textbf{(f)}) \\
\bottomrule
\end{tabular}
\caption{Mechanical and optical parameters used to describe the PhC devices of Tsvirkun \textit{et al.} All parameters have been retrieved from \cite{PC_tsvirkun_2015, PC_tsvirkun_thesis_2015} (mechanical mode labeled M1), except for $\bar{\kappa}$ and $\bar{\kappa}_e$ which have been deducted for each configurations.}
\label{tab:parameters}
\end{table}

\begin{figure}
	\centering
        \includegraphics[width=0.48\textwidth]{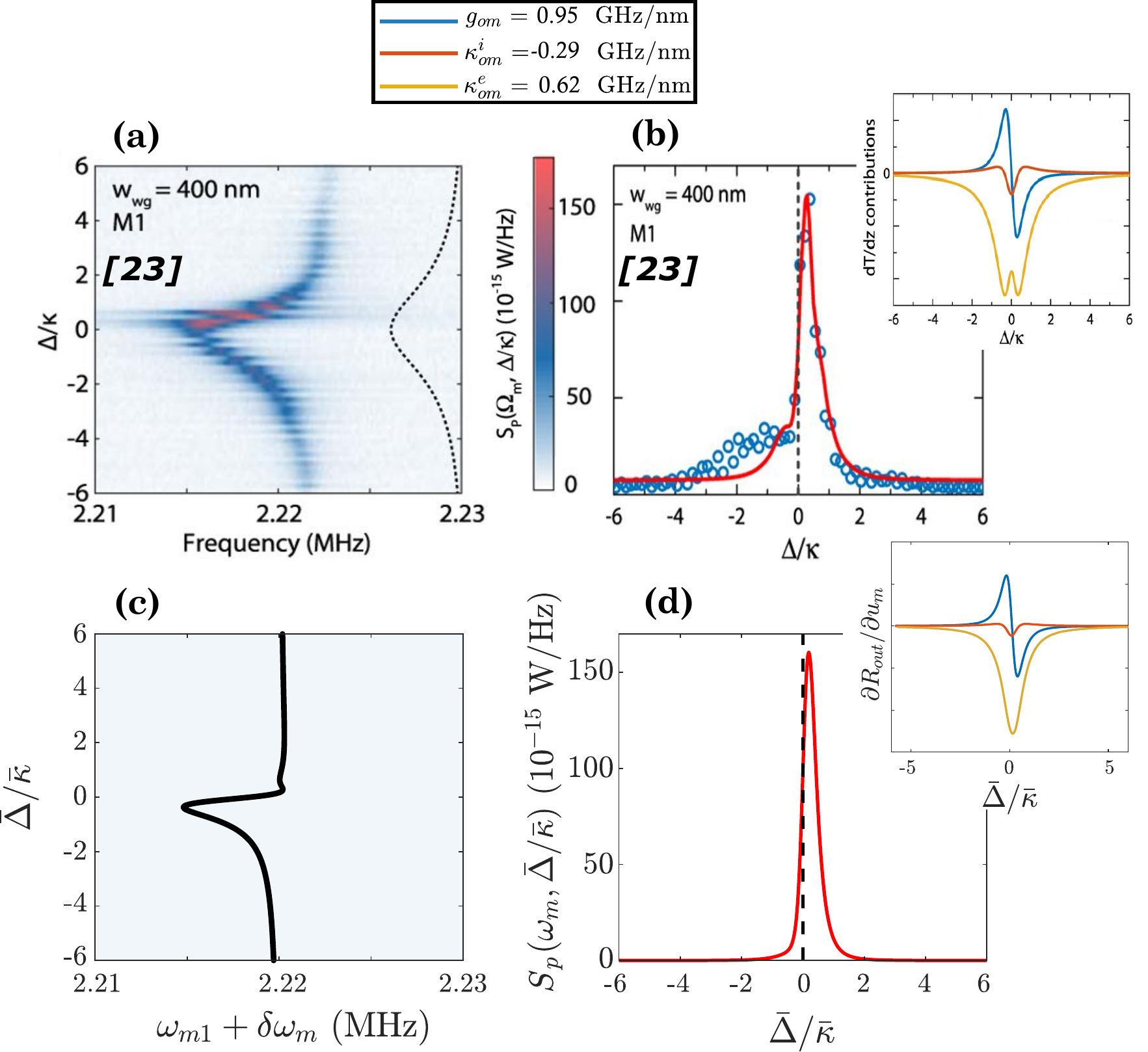} 
        \caption[ ]
        {Optical spring effect (\textbf{(a)} (from \cite{PC_tsvirkun_2015}) and \textbf{(c)} (theory)) and optical PSD at mechanical resonance frequency $S_{p}(\omega_m, \bar{\Delta})$ with $\omega_m$ depending on $\bar{\Delta}$ (\textbf{(b)} (from \cite{PC_tsvirkun_2015}) and \textbf{(d)} (theory)) as a function of the normalized optical detuning $\bar{\Delta}/\bar{\kappa}$ for an input waveguide width $w_{\textup{wg}} = \unit{450}{\nano\meter}$. The colorbar of figure \textbf{(a)} is not considered as we are only comparing the variation of the mechanical resonance frequency. The insets show the contribution of each coupling on $\deriv R_{out} / \deriv u_m$ ($g_{\textup{\textup{om}}}\partial R_{out} / \partial \bar{\Delta}$ in blue, $\kappa_{\textup{om}}^i\partial R_{out} / \partial \bar{\kappa}_i$ in red and $\kappa_{\textup{om}}^e\partial R_{out} / \partial \bar{\kappa}_e$ in yellow). Tsvirkun \textit{et al.} determined it experimentally by fiting the optical spectrum on mechanical resonance with equation \ref{eqn:output_response_oscillations}, which allows them to identify the coupling strengths.} 
        \label{fig:fig5}
\end{figure}

\begin{figure*}
	\centering
        \includegraphics[scale=0.65]{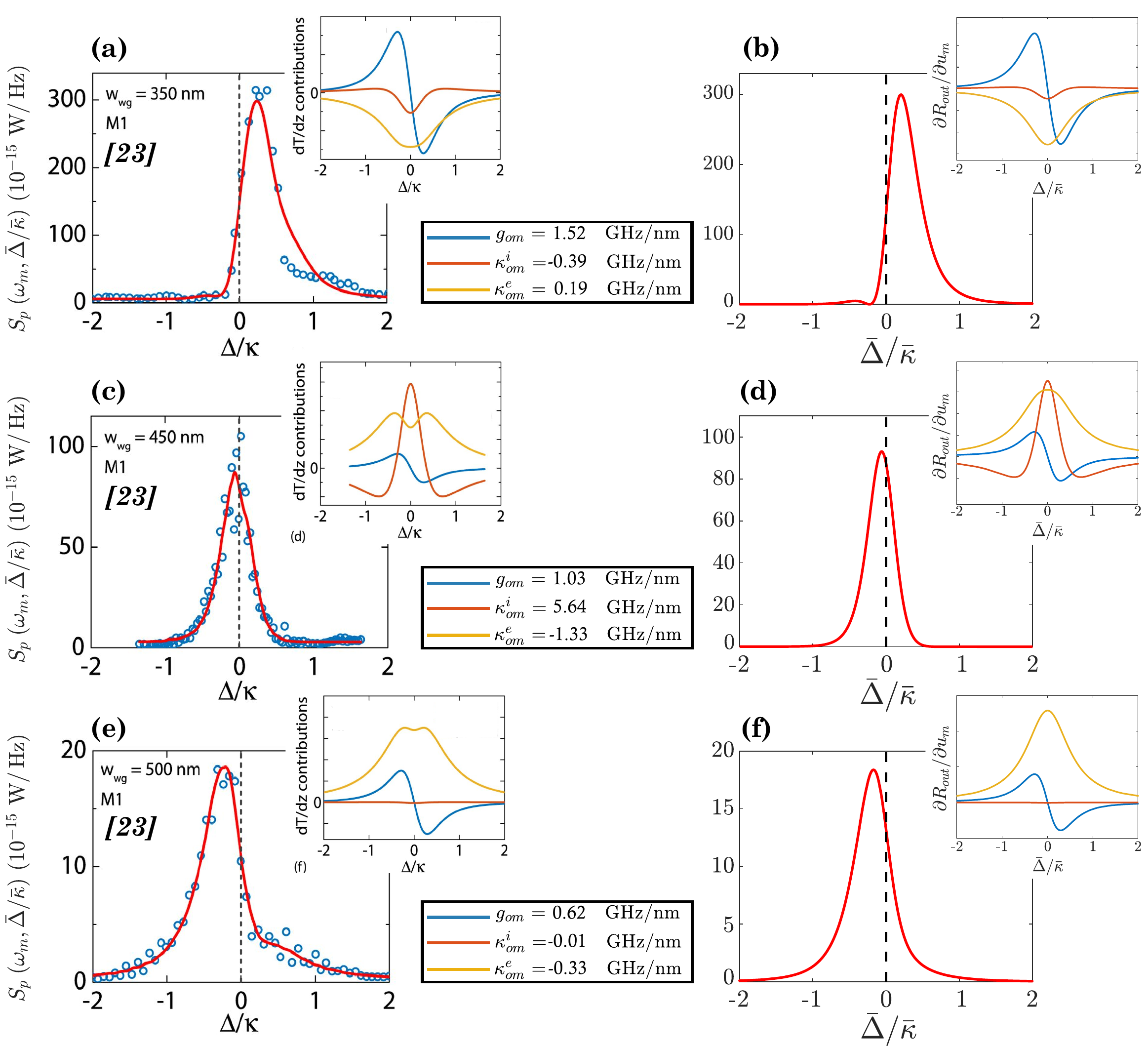} 
        \caption[ ]
        {Optical PSD at mechanical resonance frequency ($S_{p}(\omega_m, \bar{\Delta})$ with $\omega_m$ depending on $\bar{\Delta}$) as a function of the normalized optical detuning $\bar{\Delta}/\bar{\kappa}$ for different input waveguide width: $w_{\textup{wg}} = \unit{350}{\nano\meter}$ (\textbf{(a)} (from \cite{PC_tsvirkun_2015}) and \textbf{(b)} (theory)), $w_{\textup{wg}} = \unit{450}{\nano\meter}$ (\textbf{(c)} (from \cite{PC_tsvirkun_2015}) and \textbf{(d)} (theory)) and $w_{\textup{wg}}= \unit{500}{\nano\meter}$ (\textbf{(e)} (from \cite{PC_tsvirkun_2015}) and \textbf{(f)} (theory)). The insets show the contribution of each coupling on $\deriv R_{out} / \deriv u_m$ ($g_{\textup{\textup{om}}}\partial R_{out} / \partial \bar{\Delta}$ in blue, $\kappa_{\textup{om}}^i\partial R_{out} / \partial \bar{\kappa}_i$ in red and $\kappa_{\textup{om}}^e\partial R_{out} / \partial \bar{\kappa}_e$ in yellow). Note that the discrepancy on optical resonance between the insets of measurements and theory for the external dissipative contribution (\textit{i.e.} $\kappa_{\textup{om}}^e\partial R_{out} / \partial \bar{\kappa}_e$) is due to a Fano modification of the optical response in practice \cite{PC_wu_2014}.} 
        \label{fig:fig6}
\end{figure*}

We now illustrate these calculations in a concrete example in the undercoupled and unresolved sideband regimes. We consider the PhC mechanical resonator suspended over a waveguide from Tsvirkun \textit{et al.} \cite{PC_tsvirkun_2015} and compare their measurements with our own theoretical model. The parameters used to describe their devices are given in table \ref{tab:parameters}. The mechanical quality factor is chosen in the range $Q_m\sim 2000 - 3000$ \cite{PC_tsvirkun_2015, PC_tsvirkun_thesis_2015}. The optical decay rates are adjusted according to their measurements. Tsvirkun \textit{et al.} studied the same mechanical mode (labeled M1) in various configurations depending of the width $w_{\textup{wg}}$ of the input waveguide. In order to keep coherence with their measurements we take into account a constant factor in our theoretical optical spectrum and define $S_p$ in $\unit{}{\watt\per\hertz}$ as:

\begin{dmath}
S_p(\omega_m, \bar{\Delta}) = \dfrac{(\eta\beta^2g_{\textup{ti}}A)^2}{R} \, S_{\textup{opt}} (\omega_m, \bar{\Delta}) \, ,
\end{dmath}

\noindent where ``$\eta = 0.8$ is the coupling efficiency between the laser output and the lens focusing the beam onto the grating coupler, $\beta = 0.035$ is the coupling efficiency into (and out of) the access waveguide, $A = 25$ is the signal amplification, $g_{\textup{ti}} = \unit{1400}{\volt\per\watt}$ is the transimpedance gain of the photodetector and $R = \unit{50}{\ohm}$'' (see \cite{PC_tsvirkun_2015} for more details). In the following, $S_p$ is considered as the optical PSD. We made our own calculations in these systems and compared them to the measurements in figure \ref{fig:fig5} and \ref{fig:fig6}. The associated dispersive and dissipative optomechanical couplings (from their measurements) are indicated in the figures for each configuration.

The optical spring effect \textit{i.e.} the variation of the mechanical resonance frequency with the optical detuning, is compared on figure \ref{fig:fig5} (a) (from \cite{PC_tsvirkun_2015}) and (c) (analytical expression). Good agreement is found between our model and the experimental results with a similar detuning dependency and the same order of magnitude of $\unit{5}{\kilo\hertz}$ for $\delta\omega_m$. This maximum variation of the mechanical resonance frequency occurs close to optical resonance, the signature of an important dissipative behavior.

The optical PSD at mechanical resonance frequency $S_{p}(\omega_m, \bar{\Delta})$ (with $\omega_m$ depending on $\bar{\Delta}$) as a function of the normalized optical detuning is compared on figure \ref{fig:fig5} (b) (from \cite{PC_tsvirkun_2015}) and (d) (analytical expression). The insets show the contribution of each optomechanical coupling on the mechanically induced optical response oscillations (\textit{i.e.} the three terms $g_{\textup{om}}\partial R_{out} / \partial \bar{\Delta}$, $\kappa_{\textup{om}}^i\partial R_{out} / \partial \bar{\kappa}_i$ and $\kappa_{\textup{om}}^e\partial R_{out} / \partial \bar{\kappa}_e$, see equations \ref{eqn:output_response_oscillations} to \ref{eqn:output_response_Rout_derivative_diss_e}). The discrepancy on optical resonance between the insets of measurements and theory for the external dissipative contribution (\textit{i.e.} $\kappa_{\textup{om}}^e\partial R_{out} / \partial \bar{\kappa}_e$) is due to a Fano modification of the optical response in practice. We do not consider this effect as it has no impact on the optomechanical effects \cite{PC_wu_2014}. The detuning dependency is governed by the external decay rate regime \textit{i.e.} the value of $\bar{\kappa}_e$. The order of magnitude is mostly governed by the decay rates, the input power and the mechanical properties. As the last two are fixed, the decay rates are adjusted (see table \ref{tab:parameters}). The overall decay rate is kept close to $\unit{0.1 - 1}{\nano\meter}$ \cite{PC_tsvirkun_thesis_2015}. The best $\bar{\kappa}_e$ is determined by looking the contribution of each coupling on $\deriv R_{out} / \deriv u_m$ and comparing it to the measurements of Tsvirkun \textit{et al.} (see insets of figures \ref{fig:fig5} (b) and (d)). Good agreement is found in the detuning dependency of the optical PSD on mechanical resonance between measurements and theory with a single, slightly optically detuned sideband due to the comparable dispersive and external dissipative optomechanical coupling values in the undercoupled regime. The orders of magnitude of the optical PSD are comparable with a maximum close to $\unit{150}{\femto\watt\per\hertz}$.

We now compare multiple optical spectrums on mechanical resonance in figure $\ref{fig:fig6}$ depending on the input waveguide width. For each configuration, we follow the same procedure as before, and find the best $\bar{\kappa}$ and $\bar{\kappa_{e}}$ by comparing the contribution of each coupling on mechanically induced optical response oscillations in theory and in practice. For each optomechanical coupling configuration, the system is in the undercoupled regime, and the corresponding decay rates are given in table \ref{tab:parameters}. Once again, the orders of magnitude of the optical spectrums are in good agreement with the measurements, and the detuning dependencies follow the same behaviors, which validates our analytical model.

			 
\section{Conclusion}

In this paper, we have extended the theoretical framework used to describe optomechanical systems to the general case of a simultanously dispersive, external dissipative and internal dissipative coupling scheme. Although a previous theoretical study considered that dissipative coupling occurs solely with the modulation of the external optical losses and treated only the overcoupled regime \cite{TH_elste_2009}, we have higlighted, by means of a complete description of the mean optical output response, the interest of the three external decay rate regimes and the detuning dependency of the mechanically induced optical power oscillations in various optomechanical coupling configurations. The mechanical spectrum and the usual optomechanical effects (optical spring effect and optomechanical damping) have been investigated. In particular, we revealed the existence of interwined terms due to ``interferences'' between the couplings. The optical spectrum has been calculated in a concrete example and comparisons with previous measurements have shown excellent agreement. The study made in this article can be used in future optomechanical experiments to quantify the three couplings and understand their relative infuence on the optical and mechanical responses. It can also serve as a modeling tool for designing practical optomechanical devices, such as accelerometers, force sensors, gas spectroscopic, bio-photonic sensors and optical signal processing devices \cite{APPLI_metcalfe_2014}.


\appendix*

\begin{widetext}
\section{General expressions of optomechanical effects}

The general expression of each term of the optical spring effect $\delta\omega_m$ and optomechanical damping $\gamma_{\textup{opt}}$ given respectively by equations  \ref{eqn:optical_spring} and \ref{eqn:optical_damping}, are calculated here thanks to the method described in the main article. The terms due corresponding to purely dispersive, external and intrinsic dissipative situations are given in the following:

\begin{dgroup*}[noalign]
\begin{dmath}
	\delta \omega_m ^{\textup{disp}}(\omega, \bar{\Delta}) = \, \dfrac{\hbar g_{\textup{\textup{om}}}^2}{2 m_{\textup{eff}}} \, \dfrac{\bar{n}_{\textup{\textup{cav}}}}{\omega} \, Q(\omega) \, , 
\end{dmath}
\begin{dmath}
	\delta \omega_m ^{\textup{diss, e}}(\omega, \bar{\Delta}) =   \, \dfrac{\hbar (\kappa_{\textup{om}}^e)^2}{16m_{\textup{eff}}} \, \dfrac{\bar{n}_{\textup{\textup{cav}}}}{\omega\bar{\kappa}_{e0}} \, \left[ \, \dfrac{\bar{\kappa}(\bar{\kappa}/2-\bar{\kappa}_{e0}) + 2 \bar{\Delta}^2}{\bar{\kappa}_{e0}}  \, Q(\omega) -  \bar{\kappa}\bar{\Delta} \,   P(\omega) \right] \, ,
\end{dmath}
\begin{dmath}
	\delta \omega_m ^{\textup{diss, i}}(\omega, \bar{\Delta}) =  \, - \, \dfrac{\hbar(\kappa_{\textup{om}}^i)^2}{16m_{eff}} \, \dfrac{\bar{n}_{\textup{cav}}\bar{\kappa}}{\omega\bar{\kappa}_{e0}} \, \left[  \, Q(\omega) + \bar{\Delta} \,   P(\omega) \, \right]
\end{dmath}
\end{dgroup*}

\noindent and

\begin{dgroup*}[noalign]
\begin{dmath}
	\gamma_{\textup{opt}}^{\textup{disp}}(\omega, \bar{\Delta}) = \, \dfrac{\hbar g_{\textup{\textup{om}}}^2}{m_{\textup{eff}}} \, \dfrac{\bar{n}_{\textup{\textup{cav}}}}{\omega} \, \dfrac{\bar{\kappa}}{2} \, S(\omega) \, , 
\end{dmath}
\begin{dmath}
	\gamma_{\textup{opt}}^{\textup{diss, e}}(\omega, \bar{\Delta}) = \, \dfrac{\hbar (\kappa_{\textup{om}}^e)^2}{4m_{eff}} \, \dfrac{\bar{n}_{\textup{cav}}}{\omega\bar{\kappa}_{e0}}  \, \left[ \,  \dfrac{ \bar{\kappa}^2(\bar{\kappa}/2-\bar{\kappa}_{e0})+2\bar{\kappa}\bar{\Delta}^2}{4\bar{\kappa}_{e0}}  \, S(\omega)  \, + \, \bar{\Delta} \, R(\omega)  \, \right] , 
\end{dmath}
\begin{dmath}
	\gamma_{\textup{opt}}^{\textup{diss, i}}(\omega, \bar{\Delta}) =  \, \dfrac{\hbar (\kappa_{\textup{om}}^i)^2}{4m_{\textup{eff}}} \, \dfrac{\bar{n}_{\textup{\textup{cav}}}}{\omega\bar{\kappa}_{e0}}  \, \left[ \,  - \dfrac{\bar{\kappa}^2}{4} \, S(\omega) \, + \,  \bar{\Delta} \, R(\omega) \, \right] \, ,
\end{dmath}
\end{dgroup*}
\end{widetext}

\noindent where

\begin{dmath*}
\bar{n}_{\textup{cav}} = \dfrac{\bar{\kappa}_{e0}}{(\bar{\kappa}/2)^2 + \bar{\Delta}^2} \, \dfrac{P_{\textup{in}}}{\hbar \omega_L} 
\end{dmath*}

\noindent represents the steady-state intracavity photon number and $P$, $Q$, $R$, $S$ correspond to sum or difference of the Lorentzian shape effective cavity responses $|\chi_{\textup{\textup{cav}}}^{\textup{eff}}(\pm\omega)|^2$, with $\chi_{\textup{\textup{cav}}}^{\textup{eff}}(\omega) = [\bar{\kappa}/2-i(\bar{\Delta}+\omega)]^{-1}$, weighted or not with detuning terms $\Delta \pm \omega$ according to

\begin{dgroup*}[noalign]
\begin{dmath*}
P(\omega) = |\chi_{\textup{\textup{cav}}}^{\textup{eff}}(\omega)|^2  + |\chi_{\textup{\textup{cav}}}^{\textup{eff}}(-\omega)|^2 ,
\end{dmath*}
\begin{dmath*}
Q(\omega) = (\bar{\Delta}+\omega)  |\chi_{\textup{\textup{cav}}}^{\textup{eff}}(\omega)|^2  
+ (\bar{\Delta}-\omega)  |\chi_{\textup{\textup{cav}}}^{\textup{eff}}(-\omega)|^2 , 
\end{dmath*}
\begin{dmath*}
R(\omega) = (\bar{\Delta}+\omega)  |\chi_{\textup{\textup{cav}}}^{\textup{eff}}(\omega)|^2  
-  (\bar{\Delta}-\omega)  |\chi_{\textup{\textup{cav}}}^{\textup{eff}}(-\omega)|^2 ,
\end{dmath*}
\begin{dmath*}
S(\omega) = |\chi_{\textup{\textup{cav}}}^{\textup{eff}}(\omega)|^2  -  |\chi_{\textup{cav}}^{\textup{eff}}(-\omega)|^2  .
\end{dmath*}
\end{dgroup*}

Finally, the intertwined terms linked to ``interference'' between the couplings are given in the following:

\begin{widetext}
\begin{dgroup*}[noalign]
\begin{dmath}
	\delta \omega_m^{\textup{disp,diss,e}}(\omega, \bar{\Delta}) = \, \dfrac{\hbar g_{\textup{\textup{om}}}\kappa_{\textup{om}}^e}{2m_{eff}} \, \dfrac{\bar{n}_{\textup{cav}}}{\omega\bar{\kappa}_{e0}} \, \left[ - \dfrac{\bar{\kappa}\bar{\kappa}_{e0}}{4} \, P(\omega) \, + \, \bar{\Delta}  \, Q(\omega)  \right] \, ,
\end{dmath}
\begin{dmath} 
	\delta  \omega_m ^{\textup{disp,diss,i}}(\omega, \bar{\Delta}) = \, \dfrac{\hbar g_{\textup{\textup{om}}}\kappa_{\textup{om}}^i}{4m_{eff}} \, \dfrac{\bar{n}_{\textup{cav}}}{\omega\bar{\kappa}_{e0}} \, \left[ - \dfrac{\bar{\kappa}(\bar{\kappa}/2+\bar{\kappa}_{e0})}{2} \, P(\omega)  \,  + \bar{\Delta} \, Q(\omega) \, \right] \, ,
\end{dmath}
\begin{dmath}
	\delta \omega_m^{\textup{diss,e,diss,i}}(\omega, \bar{\Delta}) = \, \dfrac{\hbar \kappa_{\textup{om}}^e\kappa_{\textup{om}}^i}{8m_{eff}} \, \dfrac{\bar{n}_{\textup{cav}}}{\omega \bar{\kappa}_{e0} } \, \left[ \,  \dfrac{\bar{\kappa}(\bar{\kappa}/2-2\bar{\kappa}_{e0}) + 2\bar{\Delta}^2  }{2\bar{\kappa}_{e0}}   \, Q(\omega) \, - \, \bar{\Delta}\bar{\kappa} \, P(\omega) \, \right]  \, ,
\end{dmath}
\end{dgroup*}

\noindent and

\begin{dgroup*}[noalign]
\begin{dmath}
	\gamma_{\textup{om}}^{\textup{disp,diss,e}}(\omega, \bar{\Delta}) =  \,  \dfrac{\hbar g_{\textup{\textup{om}}}\kappa_{\textup{om}}^e}{2m_{eff}} \, \dfrac{\bar{n}_{\textup{cav}}}{\omega\bar{\kappa}_{e0}} \, \left[ \, \bar{\kappa}\bar{\Delta}  \, S(\omega) \,  +  \, \bar{\kappa}_{e0} \, R(\omega)  \, \right] \, ,
\end{dmath}
\begin{dmath}
	\gamma_{\textup{om}}^{\textup{disp,diss,i}}(\omega, \bar{\Delta}) =  \,  \dfrac{\hbar g_{\textup{\textup{om}}}\kappa_{\textup{om}}^i}{4m_{eff}} \, \dfrac{\bar{n}_{\textup{cav}}}{\omega\bar{\kappa}_{e0}} \, \left[ \,  \bar{\kappa} \bar{\Delta}  \, S(\omega)  \, + \, ( 2\bar{\kappa}_{e0} + \, \bar{\kappa} ) \, R(\omega) \, \right] \, ,
	\end{dmath}
\begin{dmath}
	\gamma_{\textup{om}}^{\textup{diss,e,diss,i}}(\omega, \bar{\Delta}) =  \, \dfrac{\hbar \kappa_{\textup{om}}^e\kappa_{\textup{om}}^i}{4m_{eff}} \, \dfrac{\bar{n}_{\textup{cav}}}{\omega\bar{\kappa}_{e0}} \, \left[ \,  \dfrac{\big( \bar{\kappa}(\bar{\kappa}/2-2\bar{\kappa}_{e0})+2\bar{\Delta}^2 \big)\bar{\kappa}}{4\bar{\kappa}_{e0}} \, S(\omega) \, + \, 2\bar{\Delta} \, R(\omega)    \, \right] \, .
\end{dmath}
\end{dgroup*}
\end{widetext}


\bibliography{biblio}

\end{document}